\documentclass{article}

\usepackage{graphicx}

\begin{document}
\title{Quantum Measurement as a Final-State Interaction with a Macroscopic External System}
\author{K.-E. Eriksson\\ Faculty of Science and Technology\\Karlstad University
SE 651 88 Karlstad, Sweden}\maketitle

\date{}
\begin{abstract}
A small quantum scattering system (the microsystem) is studied in interaction with a large system (the macrosystem) described by unknown stochastic variables. The interaction between the two systems is diagonal for the microsystem in a certain orthonormal basis, and the interaction gives an imprint on the macrosystem. Moreover, the interaction is assumed to involve only small transfers of energy and momentum between the two systems (as compared to typical energies/momenta within the microsystem). The analysis is carried out within scattering theory. Calculated in the conventional way, the transition amplitude for the whole system factorizes. The interaction taking place within the macrosystem is assumed to depend on the stochastic variables in such a way that, on the average, no particular basis vector state of the microsystem is favoured. The density matrix is studied in a formalism which includes generation of the ingoing state and absorption of the final state. Then the dependence of the final state on the conventional scattering amplitude for the microsystem is highly non-linear.

In the thermodynamic limit of the macrosystem, the density matrix of the ensemble (of microsystem plus macrosystem) develops into a final state which involves a set of macroscopically distinguishable states, each with the microsystem in one of the basis vector states and the macrosystem in an entangled state.

For an element of the ensemble, i.e., for a single measurement, the result is instead a random walk, where the microsystem ends up in one of the basis vector states (reduction of the wave packet).

Thus, the macrosystem can be interpreted as a measurement device for performing a measurement on the microsystem. The whole discussion is carried out within quantum mechanics itself without any modification or generalization.
\end{abstract}

\section{Introduction: Quantum measurement as a process to be understood within quantum mechanics}
It is often argued that the reduction of the wave function of a quantum system in connection with a measurement process cannot be understood within quantum mechanics itself due to the linear nature of the theory. We show here that if the measurement interaction is included in the quantum-mechanical description, then the amplitude of the quantum process itself---which plays the role of the wave function---enters in a non-linear way (see Eqs (33) and/or (105) below). If the measurement apparatus (which is unknown in detail) is represented by a large number of stochastic variables, then the reduction of the wave function results in the thermodynamic limit.

This paper is a revised and extended version of  a previous paper \cite{ONE}. (We refer to this paper for more references.)

A process in a microscopic quantum system $\mu$ is described together with a related interaction with a macroscopic system $A$ (the measurement apparatus), not known in any detail and therefore described by stochastic variables.

We assume that an observable $R$ with non-degenerate eigenstates $|j\rangle_{\mu}$ is to be measured. We assume the interaction between $\mu$ and $A$ to be such that the state $|j\rangle_{\mu}$ of $\mu$ makes an imprint on $A$ without the state of $\mu$ being changed. This leads to an entanglement of $A$ with $\mu$. The imprint on $A$ by $\mu$ is made with a small energy and momentum transfer. For each $j$, the interaction with $\mu$ induces $A$ to set off along a specific succession of states with an increasing number of degrees of freedom involved. The notion of metastability of $A$ will be more precisely defined in Section 3.

We use S-matrix theory, based on quantum field theory, to analyse the interaction within $\mu$ and the interaction between $\mu$ and $A$ as a whole. The resulting transition probabilities then are non-linear in terms of the transition probabilities for a pure $\mu$ process (without $A$).

Moreover, the unknown stochastic variables of $A$ are allowed to have an enhancing or inhibiting influence on the transitions within $A$ to a final state. Therefore, the different initial states of $A$, described by stochastic variables, compete on an unequal basis to reach the final state, and the ensemble of final states can have a very different composition from that of the initial states.

The system $A$ should not only be metastable; it should also be unbiased. We take this to mean that the corresponding enhancement factors and inhibition factors of $A$ occur with the same frequency in the initial state.

In the limit of low energy and momentum transfer, the $\mu-A$  interaction factorizes in the scattering amplitude (before normalization) and hence also in the transition rate. This factor from $\mu-A$ interaction depends on $\mu$ only through its final state, labelled by $j$.
 
The stochastic variables of $A$ can be introduced through a stepwise mapping procedure, thus going in steps from the situation of the microsystem $\mu$ by itself to a situation where $\mu$ interacts with a system $A$ in the thermodynamic limit, i.e., in the limit of an infinite number of stochastic variables.

Such a mapping leads to a random walk of a kind that has been suggested earlier, with the understanding that quantum mechanics may have to be abandoned for a more general theory \cite{TWO, THREE}. In this paper, we consider a process that takes place \underline{within} linear quantum mechanics itself but produces non-linearities.

Instead of a general mapping, we have chosen here a highly simplified model for the whole stochastic dynamics. In this model, the mathematics can be carried out in detail, for a single measurement as well as for an ensemble of measurements. The procedure of increasing the number of degrees of freedom of $A$ is transparent. The result is a change in the final-state distribution over the crucial variables of $A$ from a unimodal distribution to a multimodal distribution describing the different outcomes of measurement.

The non-linear dependence of the transition probabilities for the entire process (for $\mu$ \underline{and} $A$) in terms of the transition probabilities for the pure quantum process ($\mu$ \underline{without} $A$) can be explained in perturbation theory. This is most easily done in a model with sources of the incoming states and sinks of outgoing states shown in Appendix A. We use there a method due to Kinoshita and applied in a similar context by Nakanishi \cite{FOUR}, to generalise Feynman diagrams to represent the dynamics for the elements of the final state density matrix.

\section{The microsystem: quantum decay or scattering}

Let $|0\rangle_{\mu}$ be the initial state of a scattering or decay process and $|f\rangle_{\mu}$ a final state, assumed to be different from $|0\rangle_{\mu}$,
\begin{eqnarray}
_{\mu}\langle f|0\rangle_{\mu}=0.
\end{eqnarray}
%1
Then in a plane-wave basis, the scattering operator has the matrix element
\begin{eqnarray}
_{\mu}\langle f|S|0\rangle_{\mu}=\delta^{4}(P_{f}-P_{0})_{\mu}\langle f|M|0\rangle_{\mu},~~~P_{0}=(m_{0},0,0,0),
\end{eqnarray}
%2
where $_{\mu}\langle f|M|0\rangle_{\mu}$ is the scattering amplitude.

If the initial state $|0\rangle_{\mu}$ represents an unstable system of mass $m$ and $|f\rangle_{\mu}$ a state of outgoing decay products, then
 \begin{eqnarray}
 \Gamma=(2\pi)^{-1}S_{f}\delta^{4}(P_{f}-P_{0})| _{\mu}\langle f|M|0\rangle_{\mu} |^{2},
 \end{eqnarray}
 %3
is the decay rate, with $S_{f}$ denoting integration over $P_{f}$ and summation/integration over other variables of $|f\rangle_{\mu}$.

If instead $|0\rangle_{\mu}$ represents an incoming state of two colliding particles in their centre-of-mass frame with momenta
\begin{eqnarray}
\begin{array}{l}p_{1}=(\epsilon_{1},q,0,0),\\p_{2}=(\epsilon_{2},q,0,0),~~\epsilon_{j}=\sqrt{m_{j}^{2}+q^{2}},~~j=1, 2,~\mbox{and}\\\\q=\frac{1}{2}\sqrt{m_{0}^{2}-(m_{1}+m_{2})^{2}-(m_{1}-m_{2})^{2}\left( 1 - \left( \displaystyle\frac{m_{1}+m_{2}}{m_{0}} \right) ^{2} \right) },\end{array}
\end{eqnarray}
%4
then the scattering cross section (into the set of states included in the summation $S_{f}$) is
\begin{eqnarray}
\sigma=(2\pi)^{2}\frac{\epsilon_{1}\epsilon_{2}}{qm_{0}}S_{f}\delta^{4}(P_{f}-P_{0})| _{\mu}\langle f|M|0\rangle_{\mu} |^{2}.
\end{eqnarray}
%5
The density matrix for the pure initial state $|0\rangle_{\mu}$ is
\begin{eqnarray}
\rho^{(0)}=|0\rangle_{\mu\mu} \langle 0|,
\end{eqnarray}
%6
with
\begin{eqnarray}
\mbox{Tr}\rho^{(0)}=1.
\end{eqnarray}
%7
Equations (3) and (5) then take the form
\begin{eqnarray}
\begin{array}{l} \Gamma=(2\pi)^{-1}S_{f}\delta^{4}(P_{f}-P_{0})\mbox{Tr}(\rho^{(f)}M\rho^{(0)}M^{\dagger}),\\\\\sigma=(2\pi)^{2}\displaystyle\frac{\epsilon_{1}\epsilon_{2}}{qm_{0}}S_{f}\delta^{4}(P_{f}-P_{0})\mbox{Tr}(\rho^{(f)}M\rho^{(0)}M^{\dagger}),\end{array}
\end{eqnarray}
%8
where
\begin{eqnarray}
\rho^{(f)}=|f\rangle_{\mu\mu} \langle f|.
\end{eqnarray}
%9
Here $\Gamma$ and $\sigma$ are proportional to what we may call the weight of the process,
\begin{eqnarray}
w_{0}=\mbox{Tr}(M\rho^{(0)}M^{\dagger}).
\end{eqnarray}
%10
The corresponding final (decay or scattering) state is
\begin{eqnarray}
\rho^{(s)}=w_{0}^{-1}M\rho^{(0)}M^{\dagger}=\frac{M\rho^{(0)}M^{\dagger}}{\mbox{Tr}(M\rho^{(0)}M^{\dagger})};~~\mbox{Tr}\rho^{(s)}=1.
\end{eqnarray}
 %11
Thus, in general, the probabilities here, i.e., the diagonal elements of $\rho^{(s)}$ are non-linear in the diagonal elements of $M\rho^{(0)}M^{\dagger}$.

We can rewrite (8) as
\begin{eqnarray}
\begin{array}{l} \Gamma=(2\pi)^{-1}S_{f}\delta^{4}(P_{f}-P_{0})\mbox{Tr}(\rho^{(f)}\rho^{(s)}),\\\\\sigma=(2\pi)^{2}\displaystyle\frac{\epsilon_{1}\epsilon_{2}}{qm_{0}}S_{f}\delta^{4}(P_{f}-P_{0})\mbox{Tr}(\rho^{(f)}\rho^{(s)}).\end{array}
\end{eqnarray}
%12
We introduce
\begin{eqnarray}
|\psi\rangle_{\mu}=\frac{M|0\rangle_{\mu}}{\sqrt{_{\mu}\langle0|M^{\dagger}M|0\rangle_{\mu}}}=\sum_{j=1}^{n}\psi_{j}|j\rangle_{\mu},
\end{eqnarray}
%13 
where we use a basis of eigenstates of the observable $R$ (assuming non-degenerate eigenvalues),
%~~r_{j}=r_{j}{\ast}
\begin{eqnarray}
\begin{array}{l}R|j\rangle_{\mu}=r_{j}|j\rangle_{\mu},~~r_{j}=r_{j}^{\ast},~~r_{j}\neq r_{j}~\mbox{for}~j\neq k\\\\_{\mu}\langle j|k\rangle_{\mu}=\delta_{jk},~~~_{\mu}\langle j|0\rangle_{\mu}=0,\end{array}.
\end{eqnarray}
%14
Thus (13) is the final state for $\mu$ in the absence of $A$, and
\begin{eqnarray}
\psi_{j}=\frac{M_{j}}{\sqrt{\displaystyle\sum_{l=1}^{n}|M_{l}|^{2}}},~~M_{j}=~_{\mu}\langle j|M|0\rangle_{\mu}
\end{eqnarray}
%15
and
\begin{eqnarray}
_{\mu}\langle \psi|\psi\rangle_{\mu}=\sum_{j=1}^{n}|\psi_{j}|^{2}=1.
\end{eqnarray}
%16
The non-linearity as manifested in the expression for the density matrix (11) of the outgoing state is most easily explained in a formalism involving a source of the incoming state and a sink of the outgoing state. This is presented in Appendix A, where (11) appears as the result of a unitary time development. Then, using (13), we have
\begin{eqnarray}
\rho_{\mu}^{(s)}=|\psi\rangle_{\mu\mu}\langle\psi|=\sum_{j,k=1}^{n}\psi_{j}\psi_{k}^{\ast}|j\rangle_{\mu\mu}\langle k|.
\end{eqnarray}
 %17

\section{The measurement apparatus}

We next consider the system $\mu$ together with another system $A$ with many degrees of freedom. We shall use a set of discrete stochastic variables $\underline{\mathbf{e}}$ to characterize the initial state of $A$,
\begin{eqnarray}
\rho_{A}^{(0)}=|0,0;\underline{\mathbf{e}}\rangle_{AA}\langle 0,0;\underline{\mathbf{e}}|.
\end{eqnarray}
%18
In $|0,0;\underline{\mathbf{e}}\rangle_{A}$, the first zero stands for preparedness of $A$. The second zero indicates that no signal has started propagation through $A$. We shall also use two other sets of states to characterize $A$,
\begin{eqnarray}
|j,0;\underline{\mathbf{e}}\rangle_{A}~\mbox{and}~|j,x;\underline{\mathbf{e}}\rangle_{A},~~j=1,2,~...,~n;~~x=1,2,~...,~2X\gg1.
\end{eqnarray}
%19
Here $j$ indicates that in the interaction between $\mu$ and $A$, the $j$th eigenstate of $R$ has made an imprint on $A$. In the first of these states, the zero indicates that signal propagation within $A$ has not started, whereas $x$ in the second set of states, indicates the state of propagation within $A$. For $x=2X$, the signal has reached its goal in the sense that it is ready to be irreversibly recorded. (As we shall see, during this propagation from $0$ to $2X$, the collapse of the wave function takes place. The reason for having a variable $x$ taking on values in this interval is to show how a quantitative change of $x$  describes a process that involves such a qualitative change.)

We assume the stochastic variables $\underline{\mathbf{e}}$ to be defined in such a way that they are constants of motion. They are assumed not to influence the copying process from $\mu$ to $A$ but (and even decisively) the signal propagation within $A$. In our model, we label   according to this influence. We assume copying and signal propagation for the different $j$, to be totally independent processes but also not to introduce bias for any particular measurement result.

We choose $\underline{\mathbf{e}}$ as
\begin{eqnarray}
\begin{array}{l}\underline{\mathbf{e}}=(\mathbf{e}_{1},~...,~\mathbf{e}_{n})\in\Omega,\\\\\mathbf{e}_{j}=(e_{j1},~...,~e_{j(2X)}),\end{array}
\end{eqnarray}
%20
and the set of values $\Omega$ for $\underline{\mathbf{e}}$ to be
\begin{eqnarray}
e_{jx}=\pm1;~~j=1,~...,~n;~~x=1,~...,~2X.
\end{eqnarray}
%21
We have chosen an even number $2X$ here, since this will slightly simplify the model. We assume the orthonormality conditions,
\begin{eqnarray}
\begin{array}{l}_{A}\langle0,0;\underline{\mathbf{e}}|0,0;\underline{\mathbf{e}}'\rangle_{A}=\delta_{\underline{\mathbf{e}}\underline{\mathbf{e}}'};\\\\_{A}\langle0,0;\underline{\mathbf{e}}|j,x;\underline{\mathbf{e}}'\rangle_{A}=0,~~~~~~~~~~~~~~j=1,~...,~n,~~x=0,~1, 2, ..., 2X;\\\\_{A}\langle j,x;\underline{\mathbf{e}}|k,x';\underline{\mathbf{e}}'\rangle_{A}=\delta_{jk}\delta_{xx'}\delta_{\underline{\mathbf{e}}\underline{\mathbf{e}}'}\end{array}
\end{eqnarray}
%22

For measuring the observable $R$, the measuring apparatus $A$ should be classical, metastable and non-biased. We take classical and metastable to imply the following:\\
\underline{Classical}:\\
a) the apparatus $A$ can be treated semiclassically with respect to the stochastic variables $\underline{\mathbf{e}}$, in the sense that the density matrix of $A$ ((18) generalised) is diagonal in $\underline{\mathbf{e}}$ initially and remains diagonal in $\underline{\mathbf{e}}$. Moreover, $X$ should be very large; the precise meaning of this will be made clear in the model of Section 6. Niels Bohr used to emphasize the classical nature of the measuring apparatus.\\
\underline{Metastable}:\\
b) The interaction of $A$, originally in a state of preparedness, with $\mu$ in an eigenstate $|0\rangle_{\mu}$ of $R$ leads to a corresponding impact (copying) on $A$ (without changing $\mu$), involving the transition into a propagation path, specific for the value $j$ (see Section 4).\\c) the stochastic variables $\underline{\mathbf{e}}$ influence signal propagation within $A$. Propagation up to the coordinate value $x$ involves variables within $\underline{\mathbf{e}}^{(x)}$,
\begin{eqnarray}
\begin{array}{l}\underline{\mathbf{e}}^{(x)}=(\mathbf{e}_{1}^{(x)},~...,~\mathbf{e}_{n}^{(x)}),~~\underline{\mathbf{e}}^{(2X)}=\underline{\mathbf{e}},\\\\\mathbf{e}_{j}^{(x)}=(e_{j1},~...,~e_{jx}).\end{array}
\end{eqnarray}
%23
Thus $\underline{\mathbf{e}}^{(x)}$ influences amplitudes and transition rates/partial cross sections through final state interaction (propagation). For $x=2X$, we have the full process with the whole set of stochastic variables $\underline{\mathbf{e}}$ (equations (20) and (21)) involved.

The precise meaning of a non-biased $A$ will be introduced below in connection with the assumptions concerning signal propagation.

\section{Interaction between quantum system and measurement apparatus}
				
The initial state of the combined system of $\mu$ and $A$ is a product of (6) and (18),
\begin{eqnarray}
\rho_{\mu A}^{(0)}(\underline{\mathbf{e}})=\rho_{\mu}^{(0)}\otimes\rho_{A}^{(0)}(\underline{\mathbf{e}})=|0\rangle_{\mu\mu}\langle0|\otimes|0,0,\underline{\mathbf{e}}\rangle_{AA}\langle0,0,\underline{\mathbf{e}}|.
\end{eqnarray}
%24
Here we have assumed a fixed value $\underline{\mathbf{e}}$; the generalisation to a probability distribution over $\underline{\mathbf{e}}$ will be introduced below.

Interaction (scattering or decay) within $\mu$, with $A$ staying passive, then leads to the state (with   defined in (15))
\begin{eqnarray}
\begin{array}{l}\rho_{\mu A}^{(s)}(\underline{\mathbf{e}})=\rho_{\mu}^{(s)}\otimes\rho_{A}^{(0)}(\underline{\mathbf{e}})=|\psi\rangle_{\mu\mu}\langle \psi|\otimes|0,0,\underline{\mathbf{e}}\rangle_{AA}\langle0,0,\underline{\mathbf{e}}|= \\=\frac{\displaystyle\sum_{j,k=1}^{n}M_{j}M_{k}^{\ast}|j\rangle_{\mu\mu}\langle k|\otimes|0,0;\underline{\mathbf{e}}\rangle_{AA}\langle0,0;\underline{\mathbf{e}}|}{\displaystyle\sum_{l=1}^{n}|M_{l}|^{2}}= \\=\displaystyle\sum_{j,k=1}^{n}\psi_{j}\psi_{k}^{\ast}|j\rangle_{\mu\mu}\langle k|\otimes|0,0,\underline{\mathbf{e}}\rangle_{AA}\langle0,0,\underline{\mathbf{e}}|.\end{array}
\end{eqnarray}
%25
We then include the first step of the interaction between $\mu$ and $A$, the copying interaction, resulting in the change
\begin{eqnarray}
|j\rangle_{\mu}\otimes|0,0;\underline{\mathbf{e}}\rangle_{A}\Rightarrow|j\rangle_{\mu}\otimes|j,0;\underline{\mathbf{e}}\rangle_{A},
\end{eqnarray}
%26
assumed to take place similarly in each channel. Then (25) is transformed into the state
\begin{eqnarray}
\rho_{\mu A}^{(c)}(\underline{\mathbf{e}})=\sum_{j,k=1}^{n}\psi_{j}\psi_{k}^{\ast}|j\rangle_{\mu\mu}\langle k|\otimes|j,0;\underline{\mathbf{e}}\rangle_{AA}\langle k,0;\underline{\mathbf{e}}|.
\end{eqnarray}
%27
We then have to include signal propagation within $A$, involving degrees of freedom up to position $x$, let us say,
\begin{eqnarray}
|j,0;\underline{\mathbf{e}}\rangle_{A}\Rightarrow|j,x;\underline{\mathbf{e}}\rangle_{A}~,
\end{eqnarray}
%28
which depends on the stochastic parameters $\underline{\mathbf{e}}^{(x)}$ through factors  $B_{jx}(\underline{\mathbf{e}}^{(x)})$. The state at propagation position $x$, if absorption were to take place there, would be
\begin{eqnarray}
\rho_{\mu A}^{(x)}(\underline{\mathbf{e}})=\frac{\displaystyle\sum_{j,k=1}^{n}B_{jx}(\underline{\mathbf{e}}^{(x)})B_{kx}(\underline{\mathbf{e}}^{(x)})^{\ast}\psi_{j}\psi_{k}^{\ast}|j\rangle_{\mu\mu}\langle k|\otimes|j,x;\underline{\mathbf{e}}\rangle_{AA}\langle k,x;\underline{\mathbf{e}}|}{\displaystyle\sum_{l=1}^{n}|\psi_{l}|^{2}|B_{lx}(\underline{\mathbf{e}}^{(x)})|^{2}}.
\end{eqnarray}
 %29
We can think of $x$ as describing the successive involvement of new degrees of freedom into the entanglement with $\mu$. In (29), as compared to (27), $\psi_{j}$ has been replaced by
\begin{eqnarray}
B_{jx}(\underline{\mathbf{e}}^{(x)})\psi_{j},
\end{eqnarray}
%30
and a normalisation like that in (25) has been carried out. How this kind of normalization can come about in a linear theory with unitary time evolution, is discussed in Appendix A. The factorization in (30) is due to the small energy and momentum transfer in the copying process.

The outgoing particles of the quantum process are practically on their mass shells, and the influence of charged outgoing particles on $A$ is well approximated by the current density of a classical point particles emerging from a point-like scattering centre. This is discussed in Appendix B. (Clearly, this implies a restriction on the kind of apparatus that our discussion can apply to. We find it more of an advantage to be specific on this point rather than general, being confident that a generalisation can be done in a rather straightforward way.)

As we shall see, the non-linear dependence on $B_{jx}(\underline{\mathbf{e}}^{(x)})$ in (29) has very drastic consequences for $\rho_{\mu A}^{(x)}(\underline{\mathbf{e}})$.

It is important to note that the weight analogous to (10) of the process leading to the state (29) is
\begin{eqnarray}
w_{x}(\underline{\mathbf{e}}^{(x)})=\sum_{k=1}^{n}|M_{k}|^{2}|B_{kx}(\underline{\mathbf{e}}^{(x)})|^{2}=w_{0}\sum_{k=1}^{n}|\psi_{k}|^{2}|B_{kx}(\underline{\mathbf{e}}^{(x)})|^{2}.
\end{eqnarray}
%31
Starting with an ensemble of initial states, the different sets $\underline{\mathbf{e}}$ of stochastic variables compete to reach a certain propagation state, because of the different weights (31). When taking the ensemble average over (29), then the sum in (31) cancels against the denominator of (29). We shall see that the density matrix of the ensemble becomes linear in $B_{jx}B_{kx}^{\ast}$.

For $x=2X$, considered to be the goal of competitive propagation, we introduce the notation
\begin{eqnarray}
\begin{array}{l}\rho_{\mu A}^{(f)}(\underline{\mathbf{e}})=\rho_{\mu A}^{(2X)}(\underline{\mathbf{e}}),\\\\B_{j}(\underline{\mathbf{e}})=B_{j(2X)}(\underline{\mathbf{e}}).\end{array}
\end{eqnarray}
 %32
Then
\begin{eqnarray}
\rho_{\mu A}^{(f)}(\underline{\mathbf{e}})=\frac{\displaystyle\sum_{j,k=1}^{n}B_{j}(\underline{\mathbf{e}})B_{k}(\underline{\mathbf{e}})^{\ast}\psi_{j}\psi_{k}^{\ast}|j\rangle_{\mu\mu}\langle k|\otimes|j,2X;\underline{\mathbf{e}}\rangle_{AA}\langle k,2X;\underline{\mathbf{e}}|}{\displaystyle\sum_{l=1}^{n}|\psi_{l}|^{2}|B_{l}(\underline{\mathbf{e}})|^{2}}.
\end{eqnarray}
%33
The weight of this process for a given set of stochastic variables $\underline{\mathbf{e}}$ is
\begin{eqnarray}
w_{f}(\underline{\mathbf{e}})=\sum_{k=1}^{n}|M_{k}|^{2}|B_{k}(\underline{\mathbf{e}})|^{2}=w_{0}\sum_{k=1}^{n}|\psi_{k}|^{2}|B_{k}(\underline{\mathbf{e}})|^{2},
\end{eqnarray}
%34
where we have used (31).

The \underline{relative} weight for $\underline{\mathbf{e}}$ in the final state is
\begin{eqnarray}
\frac{P(\underline{\mathbf{e}})\displaystyle\sum_{k=1}^{n}|\psi_{k}|^{2}|B_{k}(\underline{\mathbf{e}})|^{2}}{\displaystyle\sum_{\underline{\mathbf{e}}'}P(\underline{\mathbf{e}}')\displaystyle\sum_{l=1}^{n}|\psi_{l}|^{2}|B_{l}(\underline{\mathbf{e}}')|^{2}},
\end{eqnarray}
%35
where $P(\underline{\mathbf{e}})$ is the probability for $\underline{\mathbf{e}}$ in the initial state. If all $\underline{\mathbf{e}}$ are equally probable, i.e., for
\begin{eqnarray}
P(\underline{\mathbf{e}})=2^{-2nX},
\end{eqnarray}
%36
the relative weight is
\begin{eqnarray}
q(\underline{\mathbf{e}})=2^{-2nX}\sum_{k=1}^{n}|\psi_{k}|^{2}|b_{k}(\underline{\mathbf{e}})|^{2},~~~\sum_{\underline{\mathbf{e}}}q(\underline{\mathbf{e}})=1,
\end{eqnarray}
%37
where
\begin{eqnarray}
b_{k}(\underline{\mathbf{e}})=\frac{B_{k}(\underline{\mathbf{e}})}{B},
\end{eqnarray}
%38
and
\begin{eqnarray}
B=\sqrt{\langle|B_{k}(\underline{\mathbf{e}})|^{2}\rangle_{\underline{\mathbf{e}}}}=\sqrt{2^{-2nX}\sum_{\underline{\mathbf{e}}}|B_{k}(\underline{\mathbf{e}})|^{2}}
\end{eqnarray}
%39
is assumed to be the same for all $k$. We shall see that this assumption is satisfied through our interpretation of the apparatus $A$. Inserted into (38), eq. (39) implies that
\begin{eqnarray}
\langle|b_{k}(\underline{\mathbf{e}})|^{2}\rangle_{\underline{\mathbf{e}}}=1,~~~k=1,~...,~n,.
\end{eqnarray}
%40
We have discussed already the metastability of $A$. We shall now specify $B_{jx}(\underline{\mathbf{e}}^{(x)})$ of (29) and the condition that $A$ is a non-biased measuring instrument. We define $B_{jx}(\underline{\mathbf{e}}^{(x)})$ through the recursive relations
\begin{eqnarray}
\begin{array}{l}B_{jx}(\underline{\mathbf{e}}^{(x)})=B_{j~x-1}(\underline{\mathbf{e}}^{(x-1)})C_{jx},\\\\C_{jx}=c_{jx}(1+\frac{1}{2}\eta_{jx}e_{jx}-\frac{1}{8}\eta_{jx}^{~~2})\displaystyle\prod_{k\neq j}c_{kx}(1-\textstyle\frac{1}{2}\displaystyle\eta_{kx}e_{kx}-\textstyle\frac{1}{8}\displaystyle\eta_{kx}^{~~2}),\\\\B_{j0}=1,\end{array}
\end{eqnarray}
%41
where
\begin{eqnarray}
\begin{array}{l}c_{jx}=c_{jx}^{\ast},~~|c_{jx}-1|\ll1;\\\\\eta_{jx}=\eta_{jx}^{\ast},~~0<\eta_{\mathrm{min}}<\eta_{jx}\ll1.\end{array}
\end{eqnarray}
%42
In (41), a positive (negative) $e_{jx}$ strengthens (weakens) the $j$th channel and weakens (strengthens) all others, because there is a mutual anticoincidence between the channels. The factors for strengthening or weakening the occurrence or non-occurrence of a certain channel for a certain propagation position $x$ are the same; both values of (21) have the same \`a priori probability. This is our understanding of the non-bias of the measuring apparatus $A$.

Using also (21) we get the following averages over the stochastic variables (to second order in the $\eta$'s),
\begin{eqnarray}
\langle C_{jx}^{~~2}\rangle_{\underline{\mathbf{e}}^{(x)}}=\prod_{l=1}^{n}c_{lx}^{~~2},
\end{eqnarray}
%43
and
\begin{eqnarray}
\langle B_{jx}^{~~2}\rangle_{\underline{\mathbf{e}}^{(x)}}=\prod_{y=1}^{x}\prod_{l=1}^{n}c_{ly}^{~~2}.
\end{eqnarray}
%44
For $x=2X$ in (44), we get from (32)
\begin{eqnarray}
\langle B_{j}^{~2}\rangle_{\underline{\mathbf{e}}}=B^{2};~~B^{2}=\prod_{x=1}^{2X}\prod_{l=1}^{n}c_{lx}^{~~2},
\end{eqnarray}
%45
which verifies (39).
 
The non-bias of $A$ is thus manifest in the sense that $C_{jx}$ and $C_{kx}$ with $k\neq j$ in (41) depend on the variables $e_{lx}$ by factors that change into each other for $e_{jx}\leftrightarrowÐe_{kx}$, and the frequencies for the two cases in the initial state are the same according to (36).

We could have introduced independent random phase factors in $C_{jx}$. We have not done so here because it is not needed. For the correlations between the $C_{jx}$, we get (to order $\eta^{2}$)
\begin{eqnarray}
\begin{array}{l}\langle C_{jx}C_{kx}\rangle_{\underline{\mathbf{e}}}=(1-\frac{1}{2}(1-\delta_{jk})(\eta_{jx}^{~~2}+\eta_{kx}^{~~2}))\displaystyle\prod_{l=1}^{n}c_{lx}^{~~2}=\\=e^{(-\frac{1}{2}(1-\delta_{jk})(\eta_{jx}^{~~2}+\eta_{kx}^{~~2})}\displaystyle\prod_{l=1}^{n}c_{lx}^{~~2}.\end{array}
\end{eqnarray}
%46
For the final state, we get to the same order
\begin{eqnarray}
\langle B_{j}B_{k}\rangle_{\underline{\mathbf{e}}}=B^{2}\mbox{exp}\left(-\textstyle{1\over2}\displaystyle(1-\delta_{jk})\sum_{x=1}^{2X}(\eta_{jx}^{~~2}+\eta_{kx}^{~~2})\right).
\end{eqnarray}
%47
This agrees with (45) for $j=k$, and goes to zero for $j\neq k$ in the limit of infinite $X$.

\section{Statistical description of measurement dynamics}

We shall now review the dynamics of the microsystem $\mu$ in interaction with the macrosystem $A$ described by the stochastic variables $\underline{\mathbf{e}}$ on an ensemble level.

We then start with the the whole ensemble of ingoing states, each state (24) entering with equal probability (36),
\begin{eqnarray}
\overline{\rho}^{(0)}=|0\rangle_{\mu\mu}\langle0|\otimes2^{-2nX}\sum_{\underline{\mathbf{e}}}|0,0;\underline{\mathbf{e}}\rangle_{AA}\langle0,0;\underline{\mathbf{e}}|.
\end{eqnarray}
%48
The scattering taking place within $\mu$, leads to the ensemble of scattering states of the type (25),
\begin{eqnarray}
\begin{array}{l}\overline{\rho}^{(s)}=2^{-2nX}\displaystyle\sum_{\underline{\mathbf{e}}}\rho_{\mu A}^{(s)}(\underline{\mathbf{e}})=\\=\displaystyle\sum_{j,k=1}^{n}\psi_{j}\psi_{k}^{\ast}|j\rangle_{\mu\mu}\langle k|\otimes2^{-2nX}\displaystyle\sum_{\underline{\mathbf{e}}}|0,0;\underline{\mathbf{e}}\rangle_{AA}\langle0,0;\underline{\mathbf{e}}|,\end{array}
\end{eqnarray}
%49
where $\psi_{j}$ is given in terms of the scattering amplitudes by (15). So far, this is a trivial extension of the dynamics of $\mu$.
 
The scattering is followed by interaction between $\mu$ and $A$, in the form of copying. The ensemble of copied states (27) is
\begin{eqnarray}
\begin{array}{l}\overline{\rho}^{(c)}=2^{-2nX}\displaystyle\sum_{\underline{\mathbf{e}}}\rho_{\mu A}^{(c)}(\underline{\mathbf{e}})=\\=\displaystyle\sum_{j,k=1}^{n}\psi_{j}\psi_{k}^{\ast}|j\rangle_{\mu\mu}\langle k|\otimes2^{-2nX}\displaystyle\sum_{\underline{\mathbf{e}}}|j,0;\underline{\mathbf{e}}\rangle_{AA}\langle k,0;\underline{\mathbf{e}}|,\end{array}.
\end{eqnarray}
%50
Here $A$ has become entangled with $\mu$. There are new non-zero components but of the same size as the corresponding states in (49). We note that whereas the restriction of $\overline{\rho}^{(s)}$ to $\mu$ is the full density matrix of $\mu$, the corresponding restriction of $\overline{\rho}^{(c)}$ is diagonal,
\begin{eqnarray}
\begin{array}{l}\mbox{Tr}_{A~}\overline{\rho}^{(s)}=\displaystyle\sum_{j,k=1}^{n}\psi_{j}\psi_{k}^{\ast}|j\rangle_{\mu\mu}\langle k|;\\\mbox{Tr}_{A~}\overline{\rho}^{(c)}=\displaystyle\sum_{j=1}^{n}|\psi_{j}|^{2}|j\rangle_{\mu\mu}\langle  j|.\end{array}
\end{eqnarray}
%51

The next set of processes is signal propagation, i.e., the increase within $A$ of the number of degrees of freedom taking part in the entanglement. We thus start from (29) and (31), noting that the relative weight for $\underline{\mathbf{e}}^{(x)}$ (i.e., the probability for $\underline{\mathbf{e}}^{(x)}$, if final absorption were to take place at the stage $x$ of propagation) is
\begin{eqnarray}
q_{x}(\underline{\mathbf{e}}^{(x)})=\frac{w_{x}(\underline{\mathbf{e}}^{(x)})}{\displaystyle\sum_{\underline{\hat{\mathbf{e}}}^{(x)}}w_{x}(\underline{\hat{\mathbf{e}}}^{(x)})}=2^{-nx}\sum_{l=1}^{n}|\psi_{l}|^{2}|b_{lx}(\underline{\mathbf{e}}^{(x)})|^{2}.
\end{eqnarray}
%52
Here we have introduced, in analogy to (39) and (38),
\begin{eqnarray}
\begin{array}{l}b_{lx}(\underline{\mathbf{e}}^{(x)})=\frac{\displaystyle B_{lx}(\underline{\mathbf{e}}^{(x)})}{\displaystyle B_{x}};~~B_{x}=\sqrt{\langle|B_{lx}(\underline{\mathbf{e}}^{(x)})|^{2}\rangle_{\underline{\mathbf{e}}^{(x)}}};\\\\\langle|b_{lx}(\underline{\mathbf{e}}^{(x)})|^{2}\rangle_{\underline{\mathbf{e}}^{(x)}}=1.\end{array}
\end{eqnarray}
%53
The ensemble of the  $x$th states of propagation (29) is then
\begin{eqnarray}
\overline{\rho}^{(x)}=2^{-n(2X-x)}\sum_{\underline{\mathbf{e}}}q_{x}(\underline{\mathbf{e}}^{(x)})\rho_{\mu A}^{(x)}(\underline{\mathbf{e}})
\end{eqnarray}
 %54
with $q_{x}(\underline{\mathbf{e}}^{(x)})$ given by (52), and
\begin{eqnarray}
\rho_{\mu A}^{(x)}(\underline{\mathbf{e}})=\frac{\displaystyle\sum_{j,k=1}^{n}b_{jx}(\underline{\mathbf{e}}^{(x)})b_{kx}(\underline{\mathbf{e}}^{(x)})^{\ast}\psi_{j}\psi_{k}^{\ast}|j\rangle_{\mu\mu}\langle k|\otimes|j,x;\underline{\mathbf{e}}\rangle_{AA}\langle k,x;\underline{\mathbf{e}}|}{\displaystyle\sum_{l=1}^{n}|\psi_{l}|^{2}|b_{lx}(\underline{\mathbf{e}}^{(x)})|^{2}}
\end{eqnarray}
%55
with the notation (53).

For  $x=2X$, we have the final ensemble (before absorption)
\begin{eqnarray}
\overline{\rho}^{(f)}=\sum_{\underline{\mathbf{e}}}q(\underline{\mathbf{e}})\rho_{\mu A}^{(f)}(\underline{\mathbf{e}}),
\end{eqnarray}
%56
with (see (37) and (33) with (38) and (39))
\begin{eqnarray}
\begin{array}{l}q(\underline{\mathbf{e}})=2^{-2nX}\displaystyle\sum_{l=1}^{n}|\psi_{l}|^{2}|b_{l}(\underline{\mathbf{e}})|^{2},\\\rho_{\mu A}^{(f)}(\underline{\mathbf{e}})=\frac{\displaystyle\sum_{j,k=1}^{n}b_{j}(\underline{\mathbf{e}})b_{k}(\underline{\mathbf{e}})^{\ast}\psi_{j}\psi_{k}^{\ast}|j\rangle_{\mu\mu}\langle k|\otimes|j,2X;\underline{\mathbf{e}}\rangle_{AA}\langle k,2X;\underline{\mathbf{e}}|}{\displaystyle\sum_{l=1}^{n}|\psi_{l}|^{2}|b_{l}(\underline{\mathbf{e}})|^{2}}.\end{array}
\end{eqnarray}
%57

The restriction of the ensemble density matrices (54) and (56) to $\mu$ is
\begin{eqnarray}
\mbox{Tr}_{A~}\overline{\rho}^{(x)}=\mbox{Tr}_{A~}\overline{\rho}^{(f)}=\sum_{j=1}^{n}|\psi_{j}|^{2}|j\rangle_{\mu\mu}\langle j|.
\end{eqnarray}
%58
The corresponding restrictions for the density matrices of the ensemble elements (55) and (57) are
\begin{eqnarray}
\begin{array}{l}\mbox{Tr}_{A~}\rho_{\mu A}^{(x)}(\underline{\mathbf{e}})=\displaystyle\sum_{j=1}^{n}p_{jx}(\underline{\mathbf{e}})|j\rangle_{\mu\mu}\langle j|,~~~p_{jx}(\underline{\mathbf{e}})=\frac{\displaystyle|b_{jx}(\underline{\mathbf{e}}^{(x)})|^{2}|\psi_{j}|^{2}}{\displaystyle\sum_{l=1}^{n}|\psi_{l}|^{2}|b_{lx}(\underline{\mathbf{e}}^{(x)})|^{2}}\\\\\mbox{Tr}_{A~}\rho_{\mu A}^{(f)}(\underline{\mathbf{e}})=\displaystyle\sum_{j=1}^{n}p_{j}^{(f)}(\underline{\mathbf{e}})|j\rangle_{\mu\mu}\langle j|,~~~p_{j}^{(f)}(\underline{\mathbf{e}})=\frac{\displaystyle|b_{j}(\underline{\mathbf{e}})|^{2}|\psi_{j}|^{2}}{\displaystyle\sum_{l=1}^{n}|\psi_{l}|^{2}|b_{l}(\underline{\mathbf{e}})|^{2}}.\end{array}
\end{eqnarray}
%59
In the next section, we shall analyze the qualitative transition taking place for $q_{x}(\underline{\mathbf{e}})$  and $\rho_{\mu A}^{(x)}(\underline{\mathbf{e}})$ defined in (52) and (55) and appearing together in (54). We shall further simplify the model described in (41) to make it analytically soluble.

\section{Simplified model for the dynamics and statistics of the quantum system in interaction with the measurement apparatus}

We simplify the model of Section 4 by putting all factors $c_{jx}$ equal to unity and by making all $\eta_{jx}$ equal. Thus, instead of (42), we have more specifically,
\begin{eqnarray}
\begin{array}{l}c_{jx}=1,~~~\eta_{jx}=\eta,~~~0<\eta\ll1,~~\mbox{for}\\\\j=1,~...,~n,~~~~x=1,~...,~2X.\end{array}
\end{eqnarray}
%60
This fixes $C_{jx}$ in (41), and $B_{jx}$ is easily determined. We go directly to the final state with $B_{j}=B_{j(2X)}$ with the result that (see (39) and (38)) $B=1$ and $b_{j}=B_{j}$. According to (38) and (39), we have
\begin{eqnarray}
|b_{j}(\underline{\mathbf{e}})|^{2}=\prod_{x=1}^{2X}[(1+\eta e_{jx})\prod_{k\neq j}(1-\eta e_{kx})].
\end{eqnarray}
%61
Also (47) is simplified,
\begin{eqnarray}
\langle b_{j}(\underline{\mathbf{e}})b_{k}(\underline{\mathbf{e}})\rangle_{\underline{\mathbf{e}}}=e^{-2X\eta^{2}(1-\delta_{jk})}.
\end{eqnarray}
%62
A short calculation using (60) gives the following recursive relation for $p_{jx}$ of (59),
\begin{eqnarray}
\begin{array}{l}p_{jx}=p_{j~x-1}+\Delta p_{jx}\\\\\Delta p_{jx}=p_{j~x-1}\displaystyle\frac{2\eta(e_{jx}-\displaystyle\sum_{l=1}^{n}p_{l~x-1}e_{lx})}{1+2\eta\displaystyle\sum_{m=1}^{n}p_{m~x-1}e_{mx}}
\end{array}
\end{eqnarray}
%63
with probability
\begin{eqnarray}
\begin{array}{l}q(e_{1x},~
...,~e_{nx})=2^{-n}\left(1+2\eta\displaystyle\sum_{l=1}^{n}p_{l~x-1}e_{lx}\right)\\\\\displaystyle\sum_{e_{1x},...,e_{nx}}q\left(e_{1x},~
...,~e_{nx}\right)=1.
\end{array}
\end{eqnarray}
%64
This can be viewed as the $x$th step of a random walk. The relevant mean values over $e_{jx}=\pm1$ are \cite{TWO}
\begin{eqnarray}
\begin{array}{l}\langle\Delta p_{jx}\rangle=0,\\\langle\Delta p_{jx}\Delta p_{kx}\rangle=4\eta^{2} p_{j~x-1}~p_{k~x-1}\left(\delta_{jk}-p_{j~x-1}-p_{k~x-1}+\displaystyle\sum_{l=1}^{n}p_{l~x-1}^{~~2}\right).
\end{array}
\end{eqnarray}
%65
They characterize the random walk, which has the corners of the probability simplex as its attractors. One way to see this is to look at the entropy
\begin{eqnarray}
S_{x}=-\sum_{j=1}^{n}p_{jx}\mbox{ln}p_{jx}
\end{eqnarray}
%66
along the random walk. For one step, we find to second order,
\begin{eqnarray}
\begin{array}{l}\Delta S_{x}=S_{x}-S_{x-1}=\displaystyle\sum_{j=1}^{n}\frac{\partial S_{x-1}}{\partial p_{j~x-1}}\Delta p_{jx}+\\\\+\frac{1}{2}\displaystyle\sum_{j,k=1}^{n}\frac{\partial^{2}S_{x-1}}{\partial p_{j~x-1}\partial p_{k~x-1}}\Delta p_{jx}\Delta p_{kx}=-\displaystyle\sum_{j=1}^{n}\left((\mbox{ln}p_{j~x-1})\Delta p_{jx}+\frac{\Delta p_{jx}^{~~2}}{2p_{j~x-1}}\right)
\end{array}
\end{eqnarray}
%67
so that
\begin{eqnarray}
\begin{array}{l}\langle\Delta S_{x}\rangle=-2\eta^{2}\displaystyle\sum_{j=1}^{n}p_{j~x-1}\left(1-2p_{j~x-1}+\displaystyle\sum_{l=1}^{n}p_{l~x-1}^{~~2}\right)=\\\\=-2\eta^{2}\displaystyle\sum_{j=1}^{n}p_{j~x-1}\left((1-p_{j~x-1})^{2}+\displaystyle\sum_{l\neq j}p_{l~x-1}^{~~2}\right)\leq0.
\end{array}
\end{eqnarray}
%68
Thus the entropy decreases until one of the corners of the probability simplex is reached. Since the expctation value of $p_{jx}$, with increasing $x$, stays at its initial value $|\psi_{j}|^{2}$, the probability of approaching the $j$th corner is $|\psi_{j}|^{2}$.

Let us now go to the ensemble of random walks (with $2X$ steps), which is a diffusion process. Then what is important in (61) and hence also in (57) and (59) is how many $e_{jx}$ are positive or negative for each  $j$. We assume $X\pm X_{j}$ cases of $e_{jx}=\pm1$. We collect the values $X_{j}$ in vector notation,
\begin{eqnarray}
\underline{X}=(X_{1},~...,~X_{n});~~~X_{j}=\frac{1}{2}\sum_{x=1}^{2X}e_{jx},~~~-X\leq X_{j}\leq X.
\end{eqnarray}
%69
There are
\begin{eqnarray}
\prod_{l=1}^{n}\frac{(2X)!}{(X+X_{j})!(X-X_{j})!}
\end{eqnarray}
%70
values of $\underline{\mathbf{e}}$ in the set $\Omega_{\underline{X}}$, characterized by (69). We then define the following $n$ distributions over $\underline{X}$,
\begin{eqnarray}
P_{j}(\underline{X})=P(X_{j})\prod_{k\neq j}P(-X_{k});~~~~\sum_{\underline{X}}P_{j}(\underline{X})=1,
\end{eqnarray}
 %71
where
\begin{eqnarray}
\begin{array}{l}P(Y)=\frac{\displaystyle(2X)!}{\displaystyle(X+Y)!(X-Y)!}\left(\frac{\displaystyle1+\eta}{\displaystyle2}\right)^{X+Y}\left(\frac{\displaystyle1-\eta}{\displaystyle2}\right)^{X-Y};\displaystyle\sum_{Y=-X}^{X}P(Y)=1,\\\\\displaystyle\sum_{Y=-X}^{X}P(Y)Y=X\eta,~~~\displaystyle\sum_{Y=-X}^{X}P(Y)Y^{2}-X^{2}\eta^{2}=\textstyle\frac{1}{2}\displaystyle X.\end{array}
\end{eqnarray}
%72
Let $Q(\underline{X})$ be the distribution over $\underline{X}$ corresponding to the distribution $q(\underline{\mathbf{e}})$ for outgoing states in (56) and (57). Then using (61), (70) and (71), we have
\begin{eqnarray}
Q(\underline{X})=\sum_{j=1}^{n}P_{j}(\underline{X})|\psi_{j}|^{2}.
\end{eqnarray}
%73
Similarly, since $p_{j}^{(f)}(\underline{\mathbf{e}})$ of (59) depends on $\underline{\mathbf{e}}$ only through $\underline{X}$, we have
\begin{eqnarray}
p_{j}(\underline{X})=\frac{\displaystyle\sum_{\underline{\mathbf{e}}\in\Omega_{\underline{X}}}q(\underline{\mathbf{e}})p_{j}^{(f)}(\underline{\mathbf{e}})}{\displaystyle\sum_{\underline{\mathbf{e}}' \in\Omega_{\underline{X}}}q(\underline{\mathbf{e}}')}=\frac{P_{j}(\underline{X})|\psi_{j}|^{2}}{\displaystyle\sum_{l=1}^{n}P_{l}(\underline{X})|\psi_{l}|^{2}}.
\end{eqnarray}
%74
For $\sqrt{X/2}\ll X \eta$, i.e., for $X \eta^{2}\gg1$, $P(Y)$ overlaps very little with $P(-Y)$, and hence for $j\neq k$, $P_{j}(X)$ and $P_{k}(X)$ defined in (71), are also almost without overlap. The result is that $Q(\underline{X})$ in (73) is multimodal.

It is convenient to change into renormalized and continuous variables,
\begin{eqnarray}
\begin{array}{l}\underline{z}=\frac{\displaystyle1}{\displaystyle\sqrt{Z}}\cdot\frac{\displaystyle\underline{X}}{\displaystyle\sqrt{X}}=X^{-\frac{3}{4}}\eta^{-\frac{1}{2}}\underline{X},\\\\Z=\eta\sqrt{X},\end{array}
\end{eqnarray}
%75
and to approximate $P(Y)$ by
\begin{eqnarray}
P(Y)=\frac{1}{\sqrt{\pi X}}e^{-(Y-X\eta)^{2}/X},
\end{eqnarray}
%76
or, with $Y=X^{\frac{3}{4}}\eta^{\frac{1}{2}}z$,
\begin{eqnarray}
P(Y)dY=\sqrt{\frac{Z}{\pi }}e^{-Z(z-\sqrt{Z})^{2}}dz.
\end{eqnarray}
 %77
This means that in $Q(\underline{X})$, $P_{j}(\underline{X})$ is narrowly centered around a specific point,
\begin{eqnarray}
Q(X)d^{n}\underline{X}=\hat{q}(\underline{z})d^{n}z,
\end{eqnarray}
%78
\begin{eqnarray}
\hat{q}(\underline{z})=\sum_{j=1}^{n}|\psi_{j}|^{2}g(\underline{z}-\underline{z}^{(j)}),
\end{eqnarray}
%79
with
\begin{eqnarray}
\begin{array}{l}g(\underline{z})=\left(\displaystyle\frac{Z}{\pi}\right)^{n/2}e^{-Z\underline{z}^{2}},\\\\z^{(j)}_{~~k}=\sqrt{Z}(2\delta_{jk}-1).\end{array}.
\end{eqnarray}
%80
For $p_{j}(\underline{X})$ in (74), only the values very close to $\underline{X}=X^{\frac{3}{4}}\eta^{\frac{1}{2}}\underline{z}^{(k)}$ are of interest, and either the numerator of (74) is negligible ($k \neq j$), or it coincides with the totally dominating term of the denominator ($k=j$). Using $\underline{z}$ as argument (rather than $\underline{X}$), we have in the limit of large $Z$,
\begin{eqnarray}
p_{j}(\underline{z}^{(k)})=\delta_{jk}.
\end{eqnarray}
%81
In this limit, the peaks of (79) separate as well as narrow down. In terms of the semi-classical stochastic variables $\underline{\mathbf{e}}$, according to (69) and (75),
\begin{eqnarray}
z_{j}=\textstyle\frac{1}{2}X^{-\frac{3}{4}}\eta^{-\frac{1}{2}}\displaystyle\sum_{x=1}^{2X}e_{jx}.
\end{eqnarray}
%82
The components of $\underline{z}$, each a sum of many small semi-classical variables, can be viewed as classical variables functioning as pointer variables. They are given by the unknown initial state of $A$, but the distribution (79) of $\underline{z}$ in the final state, depends on the interaction between $\mu$ and $A$. One can say that the components of $\underline{z}$ are (non-local) variables hidden in the unknown initial state of $A$.

The situation for $n=3$, can easily be depicted in two dimensions (Fig. 1).

Thus we have seen how the distribution $q(\underline{\mathbf{e}})$, defined in (57) and appearing in the ensemble (56) of final states, corresponds to the pointer distribution $\hat{q}(\underline{z})$ defined in (79) (with the relationship between $\underline{z}$ and $\underline{\mathbf{e}}$ given by (82)). At the $k$th peak of $\hat{q}(\underline{z})$, the corresponding state of $\mu$ is the $k$th eigenstate of the observable $R$ as indicated by (81).
 
The ensemble of final states (56) (with (57)) can be written
\begin{eqnarray}
\overline{\rho}^{(f)}=\sum_{j=1}^{n}|\psi_{j}|^{2}|j\rangle_{\mu\mu}\langle j|\otimes\overline{\rho}_{A}^{(f,j)}
\end{eqnarray}
%83
where the $n$ states of $A$,
\begin{eqnarray}
\begin{array}{l}\overline{\rho}_{A}^{(f,j)}=\left[\displaystyle\prod_{l=1}^{n}\frac{\displaystyle(2X)!}{\displaystyle(X+X^{(j)}_{~~l})!(X-X^{(j)}_{~~l})!}\right]^{-1}\times\\\\\times\displaystyle\sum_{\underline{X}}P_{j}(\underline X)\displaystyle\sum_{\underline{\mathbf{e}}\in\Omega_{\underline{X}}}|j,2X;\underline{\mathbf{e}}\rangle_{A~A}\langle j,2X;\underline{\mathbf{e}}|,\end{array}
\end{eqnarray}
%84
centered around $X^{(j)}_{~~k}=X\eta(2\delta_{jk}-1)$, are \underline{macroscopically distinguishible} due to the negligible overlap of the different $P_{j}(\underline{X})$.

The transition from a unimodal to a multimodal distribution can be followed in detail in equations (71), (72) and (73), by starting with a relatively small value for $X$ and letting it increase to a large value ($X\gg\eta^{-2}$). This means that for a moment, we have given $2X$ the role of the variable $x$ used in Sections 4 and 5.

Alternatively, one can follow this development in the continuous functions (79) and (80) (linked to (71)-(73) by (75), (77) and (78)) with increasing $Z$. The transition of $\hat{q}(\underline{z})$ from unimodal to n-modal takes place near $Z=1$.

Instead of the model developed in this section, we could have studied the consequences of the slightly more general recursive relations (41) through successive mappings along increasing $x$  in (29). These mappings are also successive steps in a random-walk or diffusion process. The mathematics would have been a bit more complicated, but the conclusion would have been of the same nature.

\section{Six comments}

\subsection{The number of detector sets}

So far, we have assumed one set of detectors for each of the $n$ eigenstates of $R$. In fact, only $n-1$ detectors are needed. If there is no detector for the $n$th eigenstate, then (41) to (45) have to be slightly modified, but (37) to (40) still hold and the conclusion, as manifested by equations (58) and (59), is the same. These leads to changes in (69), (71) and (80) in Section 6 as follows:
\begin{eqnarray}
\underline{X}=(X_{1},~...,~X_{n-1},0);~~~X_{j}=\frac{1}{2}\sum_{x=1}^{2X}e_{jx},~~~-X\leq X_{j}\leq X;
\end{eqnarray}
 %85
and
\begin{eqnarray}
\begin{array}{l}P_{j}(\underline{X})=P(X_{j})\displaystyle\prod_{k\neq j,n}P(-X_{k}),~~~j=1,~...,~n-1;\\P_{n}(\underline{X})=\displaystyle\prod_{k=1}^{n-1}P(-X_{k}),\\\\\displaystyle\sum_{\underline{X}}P_{j}(\underline{X})=1,~~~j=1,~...,~n.\end{array}
\end{eqnarray}
%86
In (80) $\underline{z}$ is now restricted to $z_{n}=0$ , and we have more precisely,
\begin{eqnarray}
\begin{array}{l}g(\underline{z})=\left(\displaystyle\frac{Z}{\pi}\right)^{n/2}e^{-Z\underline{z}^{2}},\\\\z^{(j)}_{~~k}=\sqrt{Z}(2\delta_{jk}-1);~~~j=1,~...,~n-1,~n,~~~k=1,~...,~n-1.\end{array}
\end{eqnarray}
%87
The splitting into a multimodal distribution in $\underline{z}$ is thus of the same nature as before.

\subsection{Entropy considerations}

To find the entropy of the final state, only diagonal elements of the density matrix need to be considered. For given $j$ and $\underline{X}$, there are
\begin{eqnarray}
\prod_{l=1}^{n}\frac{(2X)!}{(X+X_{l})!(X-X_{l})!}
\end{eqnarray}
%88
states, each with probability
\begin{eqnarray}
|\psi_{j}|^{2}2^{-2nX}(1-\eta^{2})^{X}\left(\frac{1+\eta}{1-\eta}\right)^{X_{j}}.
\end{eqnarray}
%89
The average over $\underline{X}$ of the logarithm of the inverse of (89) is to second order in $\eta$,
\begin{eqnarray}
\begin{array}{l}-\mbox{ln}|\psi_{j}|^{2}+2nX\mbox{ln}2-X\mbox{ln}(1-\eta^{2})-X\eta~\mbox{ln}\frac{\displaystyle1+\eta}{\displaystyle1-\eta}=\\\\=-\mbox{ln}|\psi_{j}|^{2}+X(2n\mbox{ln}2-\eta^{2}).\end{array}
\end{eqnarray}
%90
Averaging also over $j$, we find the entropy over the ensemble of final states,
\begin{eqnarray}
-\sum_{k=1}^{n}|\psi_{k}|^{2}\mbox{ln}|\psi_{k}|^{2}+2Xn\mbox{ln}2-X\eta^{2}.
\end{eqnarray}
%91
The first term here refers to the uncertainty inherent in the bifurcation following the interaction between $\mu$ and $A$, and it is not present after the bifurcation which is classical in nature (the pointer reading). The second term is an uncertainty brought in by $A$ but slightly reduced from its initial value (through correlation-building interaction, see immediately below) by the third term. Not included here is the entropy production from the final process of recording the result.

\subsection{Build-up of correlations: a microperspective}

It is easy to see, on the microlevel, the mechanism of correlation build-up for distributions of the type (52) when the number of variables is increased through increase of $x$. Let the probabilities for two different outcomes, $a$ and $b$ , be $p$ and $1-p$ , respectively. We consider only two  variables ($\varepsilon=\pm1$, $\varepsilon'=\pm1$), where positive values strengthen $a$ and negative values strengthen $b$. Then the distribution of $\varepsilon$ and $\varepsilon'$ within the final state is
\begin{eqnarray}
\begin{array}{l}\frac{1}{4}(1+\varepsilon\eta)(1+\varepsilon'\eta)p+\frac{1}{4}(1-\varepsilon\eta)(1-\varepsilon'\eta)(1-p)=\\\\=\frac{1}{4}(1+\varepsilon\varepsilon'\eta^{2})+\frac{1}{4}(\varepsilon+\varepsilon')(2p-1)\eta.\end{array}
\end{eqnarray}
%92
This gives the mean values,
\begin{eqnarray}
\langle\varepsilon\rangle=\langle\varepsilon'\rangle=(2p-1)\eta.~~~~\langle\varepsilon\varepsilon'\rangle-\langle\varepsilon\rangle\langle\varepsilon'\rangle=4p(1-p)\eta^{2}.
\end{eqnarray}
%93
Thus, the mean values of $\varepsilon$ and $\varepsilon'$ are positive for $p>\frac{1}{2}$ and negative for $p<\frac{1}{2}$. Moreover, $\varepsilon$ and $\varepsilon'$ are positively correlated.

\subsection{Orthogonality between scattering states and copied states}

Let the copying part of $A$ consist of a lattice of $N$ similar charged spin particles. We assume the spins to be originally parallel, and we assume a fast charged particle belonging to the scattering state of $\mu$ to pass along the lattice. As a result of this all the spins are rotated by a small angle $\Delta\theta$. The squared modulus of the scalar product of the new state (the state after copying) with the original state is typically
\begin{eqnarray}
\left(\mbox{cos}(\textstyle\frac{1}{2}\Delta\theta)\right)^{2N}\approx\left(1-\textstyle\frac{1}{8}\Delta\theta^{2}\right)^{2N}\approx e^{-\frac{1}{4}N\Delta\theta^{2}},
\end{eqnarray}
%94
which becomes extremely small for $N\Delta\theta^{2}\ll1$. This mechanism can thus justify the assumption
\begin{eqnarray}
|\langle0,0;\underline{\mathbf{e}}|j,0;\underline{\mathbf{e}}\rangle|^{2}=0.
\end{eqnarray}
%95
For different $j$, spins belonging to lattices in different places are rotated, and the corresponding states are mutually orthogonal, as in (22) for $x=x'=0$. This does not change for general $x$ and $x'$.

\subsection{Sketch of a Gedankenexperiment}

Let us discuss an outline of an experiment that would, in principle, simulate one step (out of very many stochastic steps) in a measurement process. The structure of such a set-up is shown in Figure 2.

A wave-packet of a spinless atom is entering at $\mathbf{A}$ and excited at $\mathbf{B}$ to a spin-1 state, then split at $\mathbf{C}$ by an inhomogenous magnetic field into $S_{z}$-components. The directions of motion of these component wave-packets are  changed so that they are all running in parallel through $\mathbf{D}$, where a magnetic field in the $z$-direction splits their energy into different levels. $\mathbf{D}$ is assumed to be long enough so that the atom returns to the ground state within $\mathbf{D}$ . The split wave-packet continues to $\mathbf{E}$ where detectors register arrival of the particle in either of the channels.

Due to the larger (smaller) phase space available for deexcitation of the component with higher (lower) energy, this component is enhanced (suppressed).

Thus, if the amplitudes for $S_{z}=-1,0,+1$ without a magnetic field in $D$, are
\begin{eqnarray}
\psi_{-},~\psi_{0},~\psi_{+}~~~~(|\psi_{-}|^{2}+|\psi_{0}|^{2}+|\psi_{+}|^{2}=1),
\end{eqnarray}
%96
then in the presence of a magnetic field $(0,0,B)$ in $D$, the probabilities for detection change from $|\psi_{-}|^{2},|\psi_{0}|^{2},|\psi_{+}|^{2})$ into
\begin{eqnarray}
\frac{(|\psi_{-}|^{2}(1-\alpha B),|\psi_{0}|^{2},|\psi_{-}|^{2}(1+\alpha B)}{1+\alpha B(|\psi_{+}|^{2}-|\psi_{-}|^{2})},
\end{eqnarray}
%97
where $\alpha$ is a constant.

Here we have assumed the decay amplitude to be constant within the relevant range of momentum for the decay photon. Then the phase-space factor is proportional to
\begin{eqnarray}
\int\frac{d^{3}\mathbf{k}}{|\mathbf{k}|}\delta(|\mathbf{k}|-...)|M(\mathbf{k})|^{2}, 
\end{eqnarray}
%98
which is proportional to $1-\alpha B$, $1$ and $1+\alpha B$, respectively, for the three cases. This is the basis of Equation (97).

It is important to note that the probability per unit time for deexcitation is proportional to
\begin{eqnarray}
1+\alpha B(|\psi_{+}|^{2}-|\psi_{-}|^{2}),
\end{eqnarray}
%99
which is the denominator of (97). Therefore, in a situation where $\varepsilon=B/|B|$ is $\pm1$ with equal frequency, the average detection probabilities are unchanged,
\begin{eqnarray}
|\psi_{-}|^{2},~|\psi_{0}|^{2},~|\psi_{+}|^{2}.
\end{eqnarray}
%100
We can think of $\varepsilon$ as simulating one stochastic variable. Then the change from (100) into
\begin{eqnarray}
\frac{(|\psi_{-}|^{2}(1-\varepsilon\alpha|B|),|\psi_{0}|^{2},|\psi_{-}|^{2}(1+\varepsilon\alpha|B|)}{1+\varepsilon\alpha|B|(|\psi_{+}|^{2}-|\psi_{-}|^{2})}
\end{eqnarray}
%101
simulates one step in the random walk. The crucial feature here, in contrast to the stochastic variables $e_{jx}$ of (21), is that $\varepsilon$ is under the control of the experimenter.

It may seem paradoxical that the components formed at $\mathbf{B}$ and split at $\mathbf{C}$ with \`a priori probabilities (100) given already at $\mathbf{B}$, can be subject to "revision" by a later interaction taking place at $\mathbf{D}$ changing the probabilities into (101).

The solution of the paradox is that in quantum mechanics, the entire process from $\mathbf{A}$ to $\mathbf{E}$ must be viewed as one whole. There is no observer intervening between $\mathbf{A}$ and $\mathbf{E}$.

Moreover, if the sign $\varepsilon$ of the magnetic field is unbiased, i.e., equally distributed between   the values $\pm1$, then the overall probabilities for detection at $\mathbf{E}$ , are those of (100), given already at $\mathbf{B}$.

In this sense the Gedankenexperiment sketched here can be viewed as simulating one step in the measurement process as modelled in Section 6 above.

\subsection{Quantum field theory and understanding measurement}

One manifestation of the Bohr-Einstein debate on quantum measurement is the famous 1935 article by Einstein, Podolski and Rosen \cite{FIVE} and Bohr's answer \cite{SIX} to that. At the time of this debate, relativistic quantum field theory had not yet been developed. The ordering theorem, i.e., the expression of time-ordered products of quantum fields in terms of normal-ordered products \cite{SEVEN}, a basis of Feynman diagrams, was not yet available.

In the tradition that followed, the measurement problem was not stated in field-theoretical terms, but stayed within first-quantized theory. Similarly, quantum field theory was used to derive the measurable features of microworld processes, but not to describe the interface between microworld and macroworld. When the infrared divergences of quantum electrodynamics appeared, one had to let the final state include also soft bremsstrahlung \cite{EIGHT}. Then the details of the final state had to depend on the experimental resolution. In this work, one was quite close to a description taking into account also the measurement apparatus, but it did not really happen.\\

%{\Large \bf Acknowledgements
\section*{Acknowledgements}
I thank Professor Kazimierz Rzazewski for discussions and for reading and criticizing an earlier version of this manuscripts. For discussions and technical support I also thank Anders Eriksson, Jens Fjelstad, Maria Grahn, Martin Nilsson-Jacobi, Kristian Lindgren and Carina Rehnstrom.\\

\appendix
\section{Scattering process with sources and sinks}

To describe a scattering process
\begin{eqnarray}
A+B\to 
C_{1}+~...+~C_{m},
\end{eqnarray}
%102
we first consider two sources emitting the incoming particles at time $-T$. We can think of them as one bilocal source creating the state $|0\rangle_{\mu}$ of incoming particles $A$ and $B$. Let us similarly consider a set of multiple sinks ready to absorb and identify the states $|1\rangle_{\mu},~...,~|n\rangle_{\mu}$ of outgoing particles $C_{1},~...,~C_{m
}$ at time $T$.

We assume the dynamics to be described by renormalized quantum field theory, where an S-matrix element can be represented by a set of connected Feynman diagrams. Here we represent the whole sum over such diagrams by a shaded circle with ingoing and outgoing lines (Fig. 3).

We shall combine such diagrams with open half-circles, marked with the corresponding states, representing the source at time $-T$ and the sinks at time $T$ of the scattering states $|j\rangle_{\mu}$ with the curved side as the active side, labelled by the the emitted or absorbed state. However, rather than the transition amplitudes, we shall describe the density matrix of the outgoing state at time $T$. 

To get the density matrix, we connect the initial (time $-T$) state $|0\rangle_{\mu}$ going into scattering, described by the scattering operator $S$, whereas the adjoint state is taken into the scattering state by $S^{\dagger}$. (This is a type of description used long ago by Kinoshita and Nakanishi. Since interaction Hamiltonians are hermitean and particle propagators are symmetric under time reversal, we get a whole series of diagrams involving emission, scattering and absorption and the inverse processes (Fig. 4)).

Taking together all diagrams, we find a geometrical series. The result can be viewed as the insertions of Fig. 5a and interpreted as renormalization of the emission process for the incoming state, and, similarly, renormalization of the process of nothing at all happening (Fig. 5b). The final-state density matrix is
\begin{eqnarray}
\begin{array}{l}
F_{j}M_{j0}J_{0}J_{0}^{\ast}M_{k0}^{\ast}F_{k}^{\ast}-F_{j}M_{j0}J_{0}\left(\displaystyle\sum_{l=1}^{n}J_{0}^{\ast}M_{l0}^{\ast}F_{l}^{\ast}F_{l}M_{l0}J_{0}\right)J_{0}^{\ast}M_{k0}^{\ast}F_{k}^{\ast}+\\\\+F_{j}M_{j0}J_{0}\left(\displaystyle\sum_{l=1}^{n}J_{0}^{\ast}M_{l0}^{\ast}F_{l}^{\ast}F_{l}M_{l0}J_{0}\right)^{2}J_{0}^{\ast}M_{k0}^{\ast}F_{k}^{\ast}+-
...=\\\\=F_{j}M_{j0}J_{0}\frac{\displaystyle1}{\displaystyle1+\sum_{l=1}^{n}J_{0}^{\ast}M_{l0}^{\ast}F_{l}^{\ast}F_{l}M_{l0}J_{0}}J_{0}^{\ast}M_{k0}^{\ast}F_{k}^{\ast}=\\\\=\frac{\displaystyle|J_{0}|^{2}F_{j}F_{k}^{\ast}M_{j0}M_{k0}^{\ast}}{\displaystyle1+|J_{0}|^{2}\sum_{l=1}^{n}|F_{l}|^{2}|M_{l0}|^{2}},
\end{array}
\end{eqnarray}
%103
where
\begin{eqnarray}
J_{0},~F_{1},~F_{2},~...,~F_{n}
\end{eqnarray}
%104
represent the source and the assembley of sinks, respectively.

The term '$1$' in the denominator of the last expression in (103) represents the case of nothing happening. With a sufficiently strong source, it can be safely neglected. Then the source contributes identical factors in numerator and denominator, and the result reduces to
\begin{eqnarray}
\frac{\displaystyle F_{j}F_{k}^{\ast}M_{j0}M_{k0}^{\ast}}{\displaystyle\sum_{l=1}^{n}|F_{l}|^{2}|M_{l0}|^{2}}.
\end{eqnarray}
%105
This depends on the absorption efficiencies of the sinks. Assuming for a moment these factors to be equal, we have again the same factors in numerator and denominator. The result is the normalized final state density matrix
\begin{eqnarray}
\frac{\displaystyle M_{j0}M_{k0}^{\ast}}{\displaystyle\sum_{l=1}^{n}|M_{l0}|^{2}},
\end{eqnarray}
%106
which was our starting point in (11) or (17). Thus the non-linearity of (11) has been explained.The matrix elements $M_{j0}$ are related to $\psi_{j}$ through (13).

The interpretation of (106) is that it describes the case without a measurement apparatus. For a realistic measurement apparatus, the factors $F_{j}$ of (105) are in general different and unknown, except for the restriction that the statistical distribution over them should not introduce any bias. Thus, we can identify them with $B_{j}(\underline{\mathbf{e}})$ of Section 4.

\section{Factorization of final state interaction}

We think of the interaction between the quantum system $\mu$ and the measurement apparatus $A$ as an electromagnetic interaction with very small energy and momentum transfer. Thus it can be described in terms of an exchange of soft photons. Emission and exchange of soft photons is an old and well-known example of factorizable processes in quantum electrodynamics. When it became understood, the picture of scattering became drastically changed, in the sense that no non-forward scattering takes place without soft-photon emission. Later, this was identified as coherent radiation from classical charged point sources moving into and out from a point-like scattering centre.
 
To show the factorization of soft photon emission and exchange, we consider an outgoing electron (charge $-e$, mass $m$) with final momentum $p$, described by a spinor $\overline{u}(p)$,
\begin{eqnarray}
p^{2}=m^{2};~~~ \bar{u}(p)(ip\cdot\gamma-m)=0,
\end{eqnarray}
%107
after emitting two soft photons with momenta $k_{1}$, $k_{2}$ and polarizations $\tau_{1}$, $\tau_{2}$,
\begin{eqnarray}
\begin{array}{l}k_{1}^{2}=k_{2}^{2}=0;~~~k_{1}\cdot\tau_{1}=k_{2}\cdot\tau_{2}=0;\\
|\mathbf{k}_{1}|,|\mathbf{k}_{2}|\ll m.\end{array}
\end{eqnarray}
%108
In the evaluation of the Feynman diagram of Fig. 5, the spinor $\overline{u}(p)$ for the outgoing electron is then replaced by an expression
\begin{eqnarray}
\begin{array}{l}e^{2}\bar{u}(p)\left[\tau_{1}\cdot\gamma\frac{\displaystyle i(p+k_{1})\cdot\gamma+m}{\displaystyle(p+k_{1})^{2}+m^{2}}\tau_{2}\cdot\gamma+(1\leftrightarrow
2)\right]\frac{\displaystyle i(p+k_{1}+k_{2})\cdot\gamma+m}{\displaystyle(p+k_{1}+k_{2})^{2}+m^{2}}=\\\\=e^{2}\frac{\displaystyle1}{\displaystyle2(p\cdot k_{1}+p\cdot k_{2})}~\bar{u}(p)\times\\\\\times\left[\frac{\displaystyle \tau_{1}\cdot\gamma(ip\cdot\gamma+m)\tau_{2}\cdot\gamma(ip\cdot\gamma+m)}{\displaystyle2p\cdot k_{1}}+(1\leftrightarrow
2)\right]=\\\\=e^{2}\frac{\displaystyle-p\cdot\tau_{1}p\cdot\tau_{2}}{\displaystyle p\cdot k_{1}+p\cdot k_{2}}\left(\frac{\displaystyle1}{\displaystyle p\cdot k_{1}}+\frac{\displaystyle1}{\displaystyle p\cdot k_{2}}\right)\bar{u}(p)=(s(k_{1})\cdot\tau_{1})(s(k_{2})\cdot\tau_{2})\bar{u}(p),\end{array}
\end{eqnarray}
%109
where
\begin{eqnarray}
s_{\mu}(k)=-e \frac{ip_{\mu}}{p\cdot k}=-e\int_{0}^{\infty}dt\int d^{3}\mathbf{x}e^{i\left(\mathbf{k}\cdot\mathbf{x}-|\mathbf{k}|t\right)}~\delta^{3}\left(\mathbf{x}-\frac{\mathbf{p}}{p_{0}}t\right)\frac{p_{\mu}}{p_{0}}
\end{eqnarray}
%110
is the Fourier transform of the current of a classical point charge $-e$ moving from $\mathbf{x}=\mathbf{0}$ at time zero with the velocity $\mathbf{p}/p_{0}$. The rest of the diagram is unchanged in the limit of small $k_{1}$, $k_{2}$. Equation (109) states that the emission of the two photons is described by one scalar emission factor for each photon. The corresponding holds for two photons being absorbed by an electron, as well as for one emitted photon and one absorbed.

For $r$ photons, use can be made of the identity
\begin{eqnarray}
\sum_{(i_{1}i_{2}...i_{m})}\frac{1}{a_{i_{1}}\left(a_{i_{1}}+a_{i_{2}}\right)...\left(a_{i_{1}}+a_{i_{2}}+...+a_{i_{m}}\right)}=\frac{1}{a_{1}a_{2}...a_{m}}.
\end{eqnarray}
%111
There is also a factor $(r!)^{-1}$. Summation over photon states and over $r$ gives rise to a coherent state generated by the classical current (110).

\bibliography{references}
\bibliographystyle{plain}

\newpage
\pagestyle{empty}

\begin{figure}[h]
\vspace{-5cm}
\centering
\includegraphics[scale=0.7]{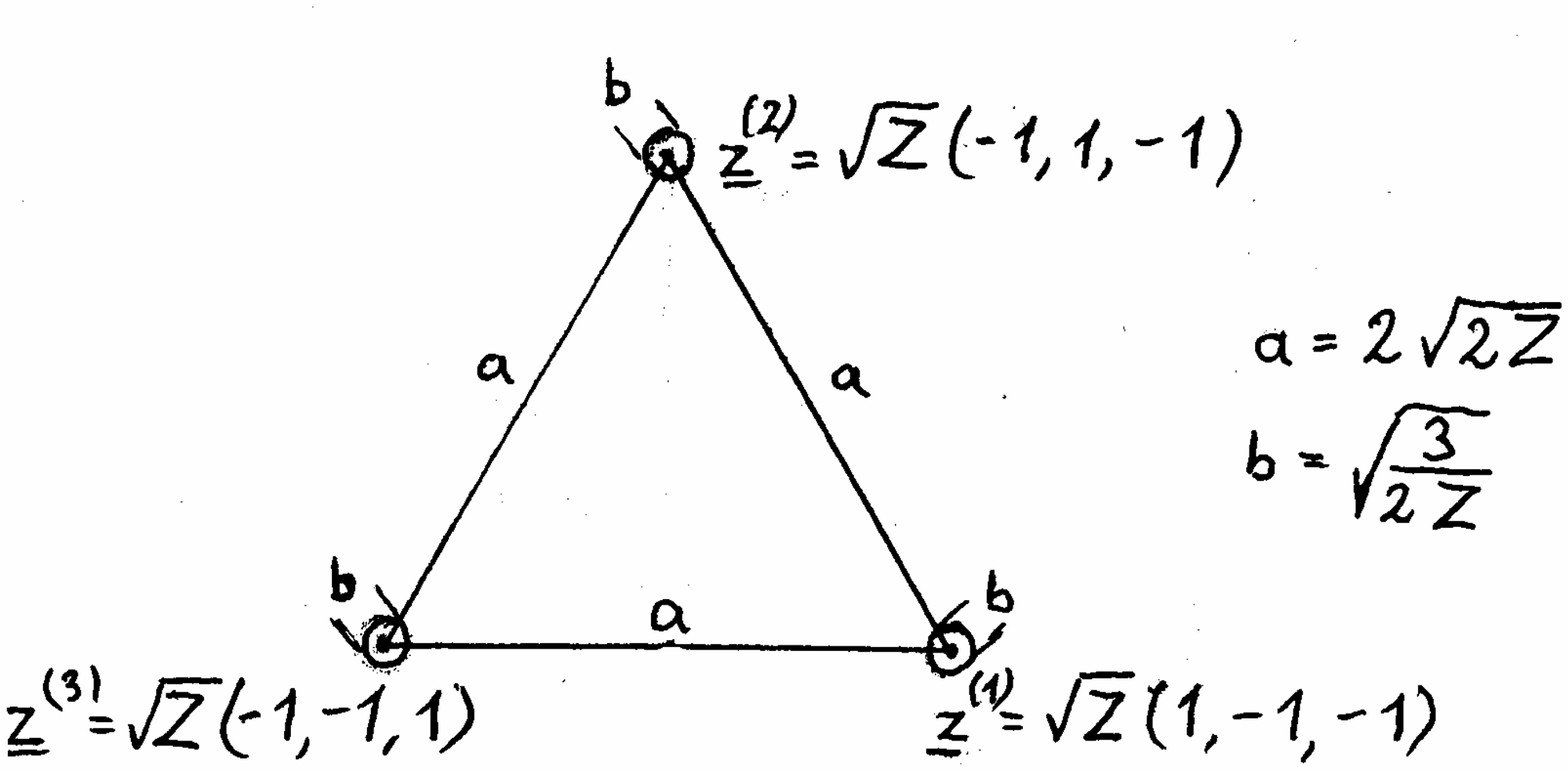}
\caption{The distribution over pointer variables $\mathbf{z}=(z_{1},z_{2},z_{3})$ for a three-state system is centered around $\mathbf{z}^{(1)}$, $\mathbf{z}^{(2)}$ and $\mathbf{z}^{(3)}$ in the plane $z_{1}+z_{2}+z_{3}+\sqrt{Z}=0$ with relative weights $|\psi_{1}|^{2}$, $|\psi_{2}|^{2}$ and $|\psi_{3}|^{2}$. For $b\ll a$, i.e., for large $Z$, the partial distributions are well separated. With increasing $Z$, the transition from a unimodal to a trimodal distribution takes place around $Z\approx1$.}
\end{figure}

\begin{figure}[h]
\vspace{-5cm}
\centering
\includegraphics[scale=0.7]{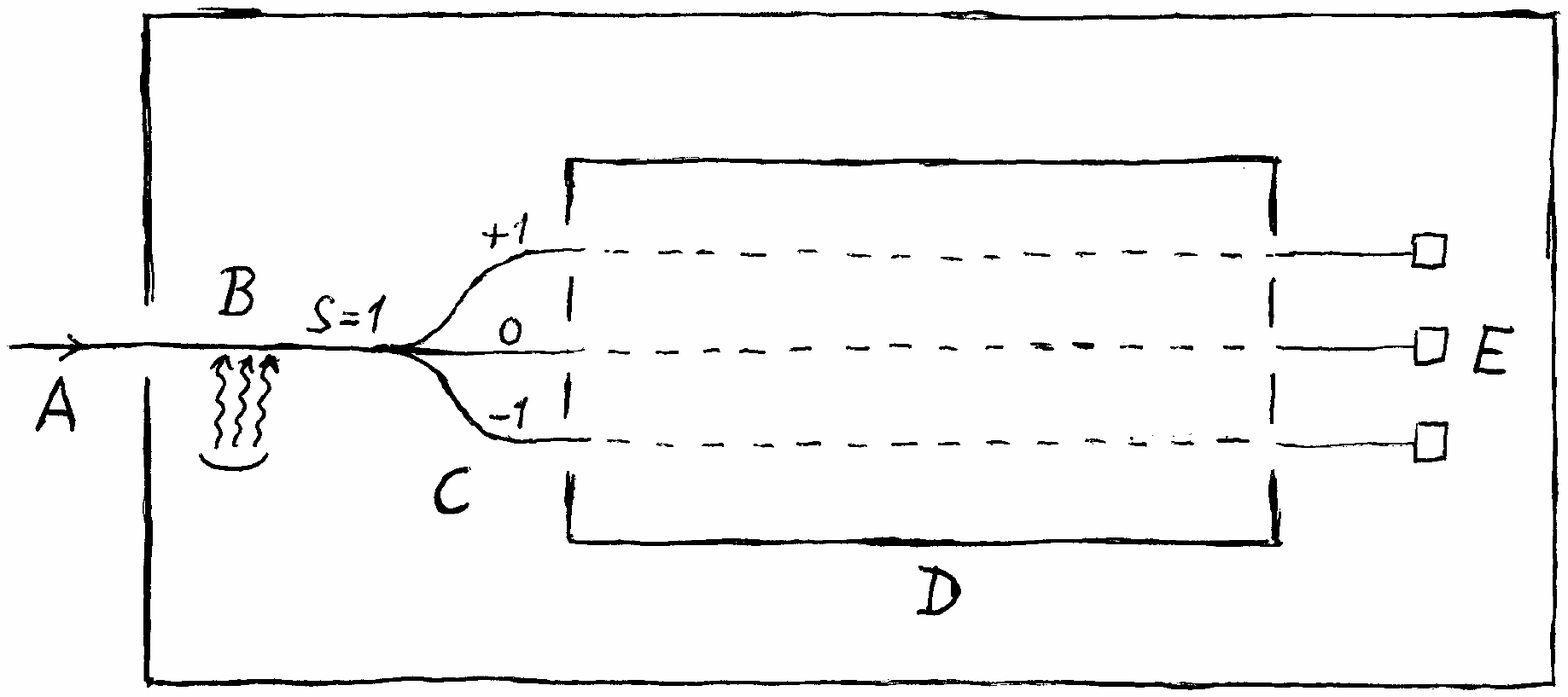}
\caption{Sketch of a Gedankenexperiment to simulate the first step in a measurement process, as described in subsection 7.5. An atomic beam is entering at A and excited by radiation at B into a spin-1 state, then split at C into the different z-components of the spin. In D there is a homogenous magnetic field $(0,0,B)$, which makes the available phase space in the relaxation to the ground state of the three $S_{z}$-components different. After deexcitation in D, the atoms are detected at E.}
\end{figure}

\begin{figure}[h]
\vspace{-5cm}
\centering
\includegraphics[scale=0.7]{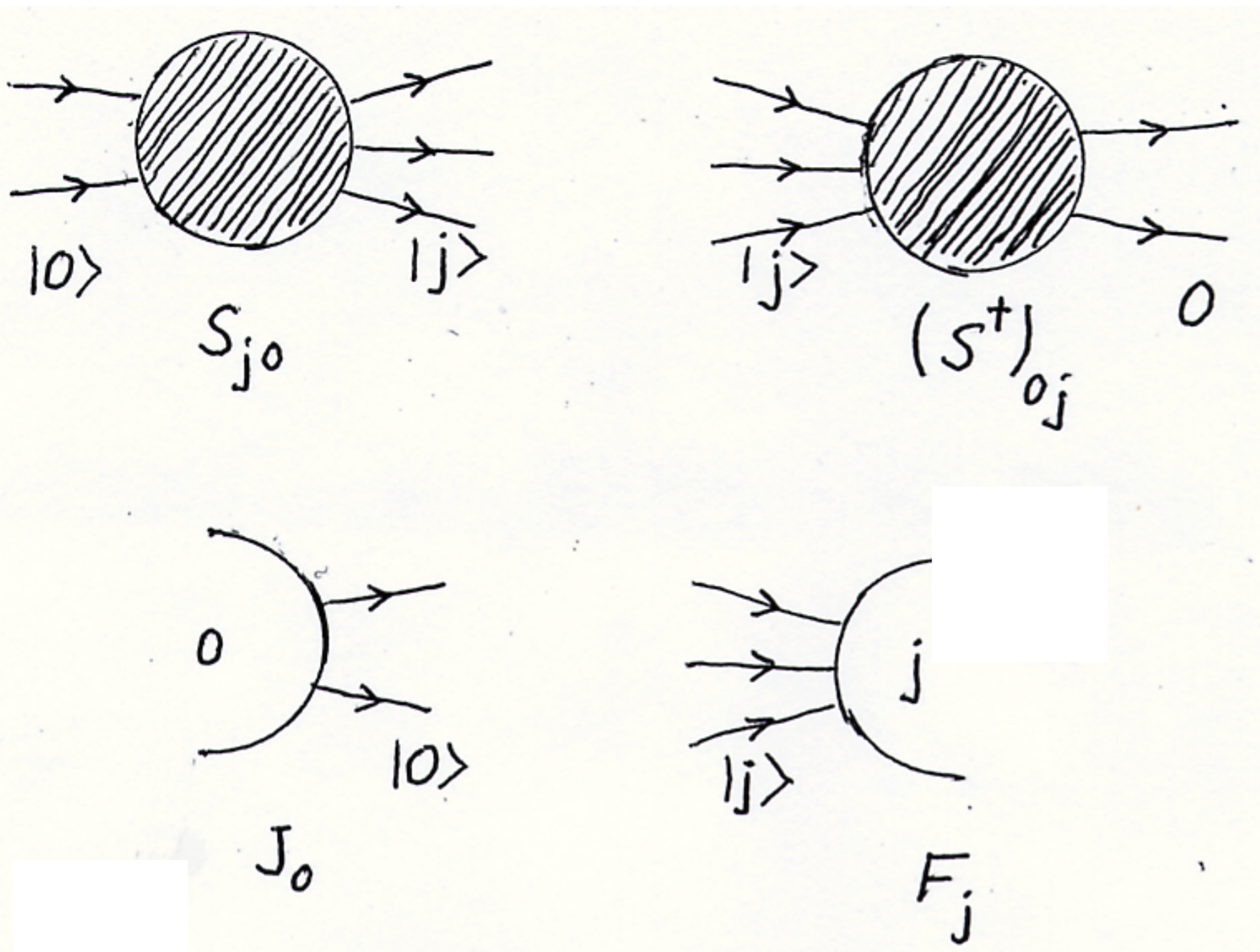}
\caption{Diagram elements for considering scattering, with production of initial particle state $|0\rangle_{\mu}$ and absorption of final particle state $|j\rangle_{\mu}$ included in the description.}
\end{figure}

\begin{figure}[h]
\vspace{-5cm}
\centering
\includegraphics[scale=0.7]{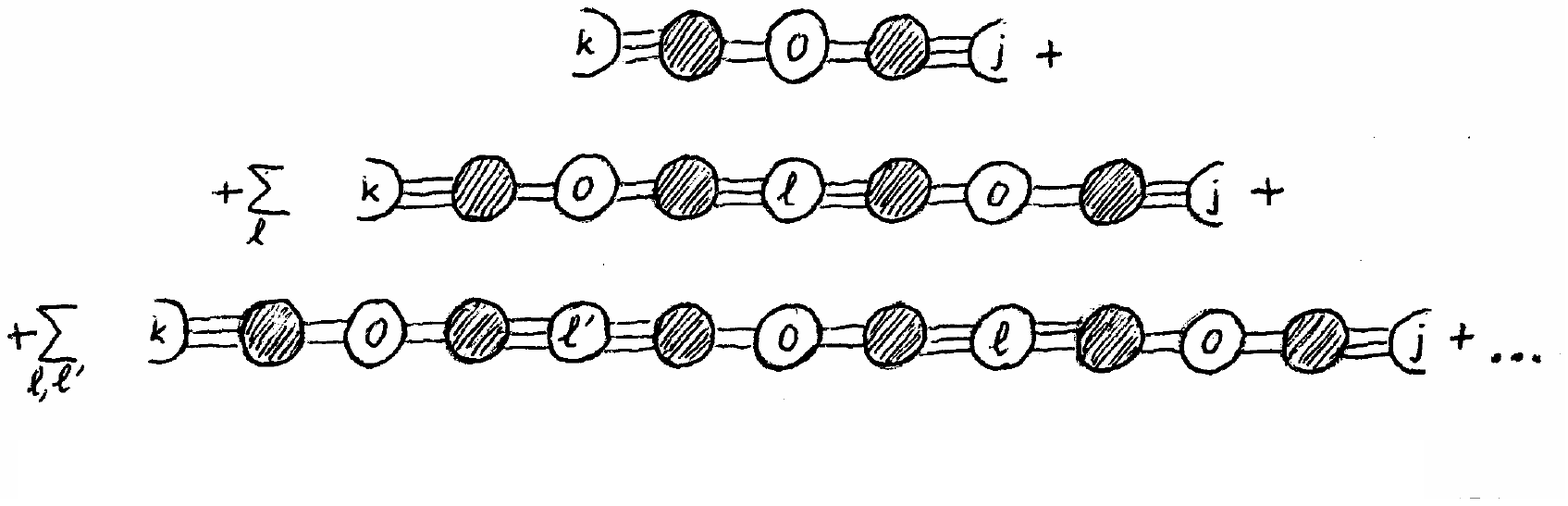}
\caption{Diagrams for the $jk$-component of the final-state density matrix, $\rho^{(f)}_{jk}$.}
\end{figure}

\begin{figure}[h]
\vspace{-5cm}
\centering
\includegraphics[scale=0.7]{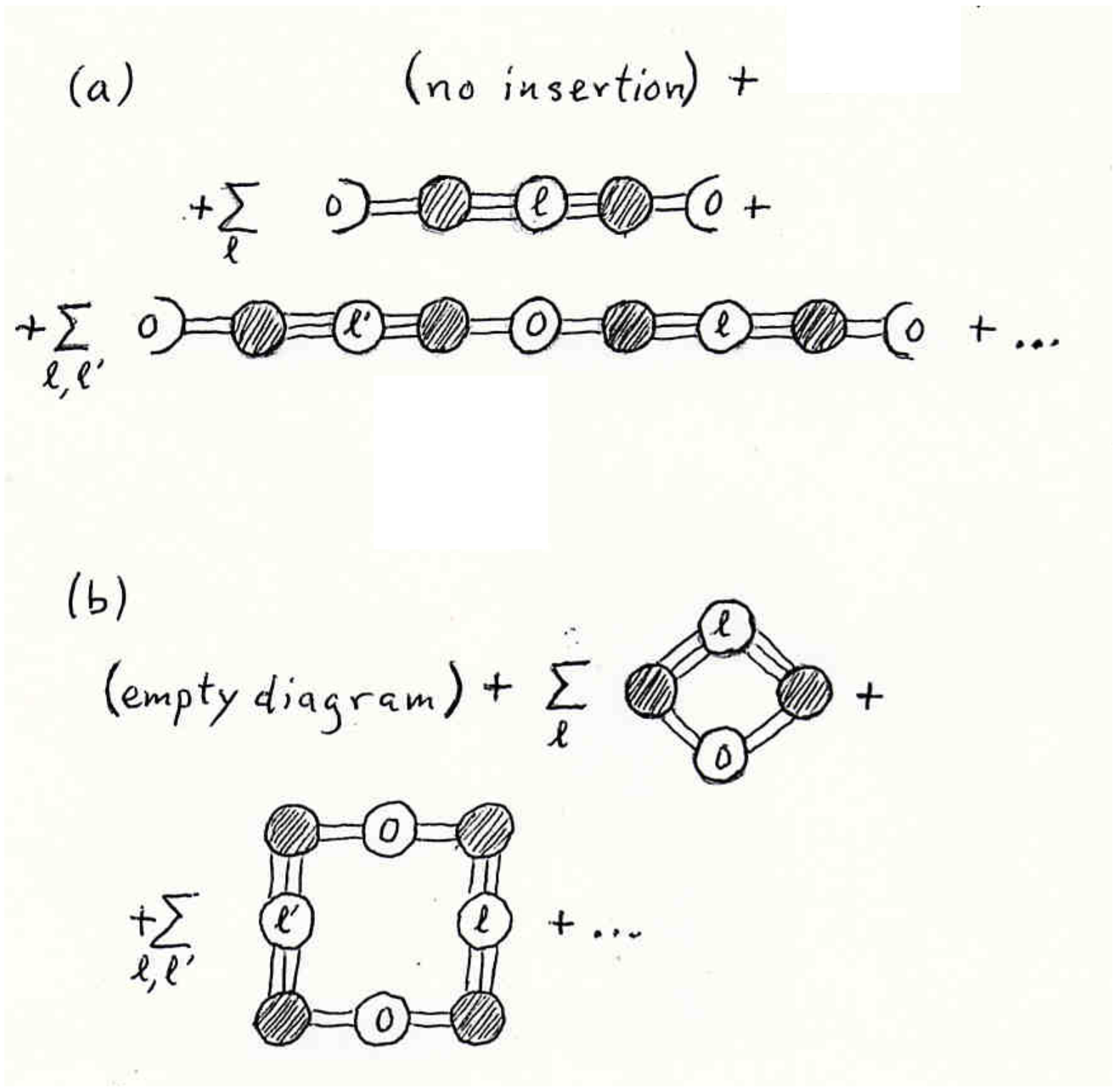}
\caption{(a) Insertions for the initial emission process into density-matrix diagrams. (b) Diagrams for the case of no process.}
\end{figure}

\begin{figure}[h]
\vspace{-5cm}
\centering
\includegraphics[scale=0.7]{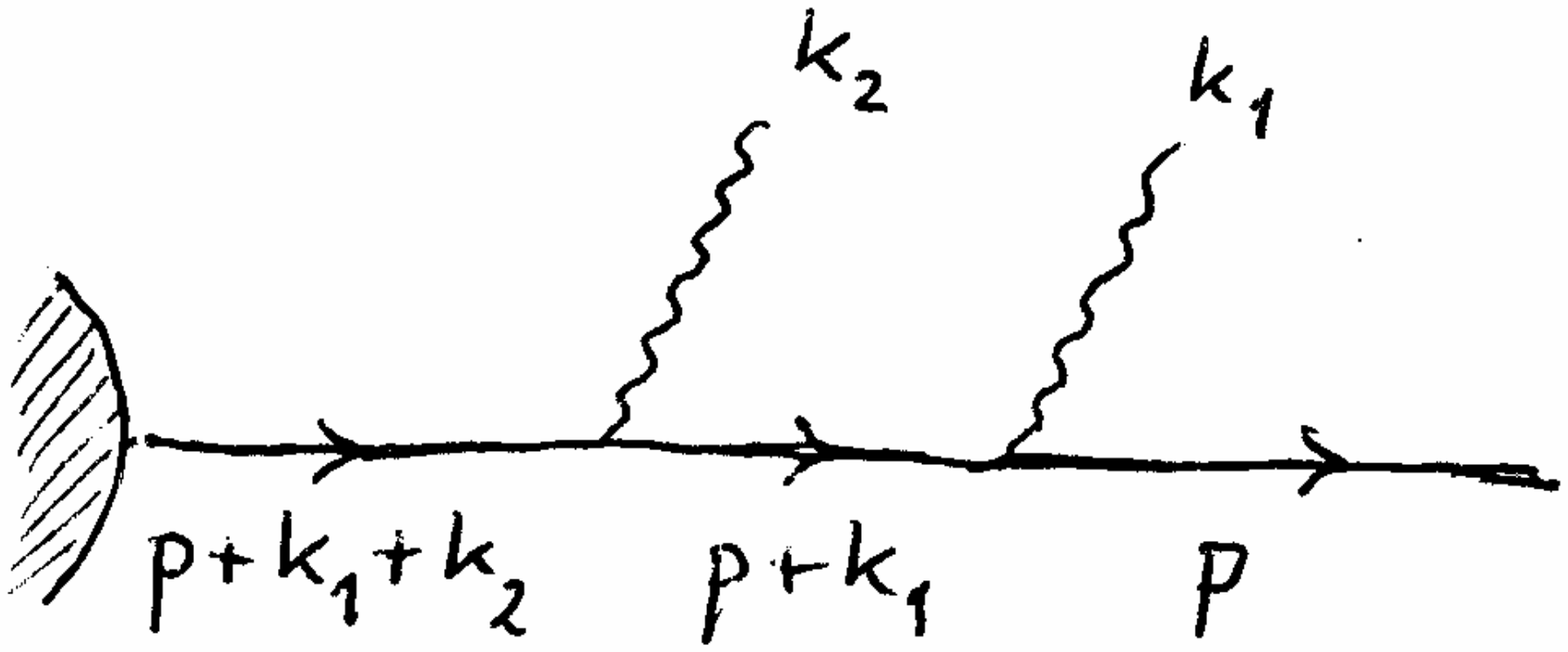}
\caption{Feynman diagram for the emission of two soft photons.}
\end{figure}

\end{document}